\documentclass[prd,aps,nofootinbib,onecolumn,showpacs,10pt]{revtex4}
\usepackage{graphicx,epsfig}

\usepackage{epsf}
\usepackage{graphicx}

\begin{document}
%\begin{fmffile}{samplepics}

\title{  \bf Thermal Spectrum of Heavy Vector and Axial Vector Mesons in the Framework of QCD Sum Rules Method }
\author{Enis Yazici \\
Turkey \\
e-mail:yazicienis@gmail.com\\}

\begin{abstract}

The masses and the leptonic decay constants of vector and axial vector heavy-heavy mesons are calculated using the thermal QCD sum rules approach. While obtaining the QCD sum rules, additional operators in the Wilson expansion and also temperature dependency of the continuum threshold are taken into account. The masses and the decay constants remained unchanged up to $T\simeq100~MeV$. After that point, they start to diminish. At the critical temperature, the masses decreased about $3\%$, $5\%$ and $14\%$ for  the vector mesons $\Upsilon$, $B_{c}$ and $J/\psi$; $6\%$, $7\%$ and $22\%$ for the axial vector mesons $\chi_{b1}$, $B_{c}$ and $\chi_{c1}$, respectively. The decay constants reached about less than $20\%$ of their vacuum values. The obtained results of the thermal mass and decay constant calculations at zero temperature are in a very good agreement with the other non-perturbative calculations at vacuum as well as with the experimental data.

\end{abstract}

\maketitle

%%%%%%%%%%%%%%%%%%%%%%%%%%%%%%%%%%%%%%%%%%%%%%%%%%%%%%%%%%%%%%%%%%%%
%\section{Introduction}
%%%%%%%%%%%%%%%%%%%%%%%%%%%%%%%%%%%%%%%%%%%%%%%%%%%%%%%%%%%%%%%%%%%%
\section{Introduction}

The heavy nuclei collisions at Large Hadron Collider (LHC) and the Relativistic Heavy Ion Collider (RHIC) experiments investigate the region of the QCD phase diagram at high temperatures. Especially, after observing $J/\psi$ suppression in these heavy ion collision experiments, this is considered as a signal of the Quark-Gluon Plasma (QGP) \cite{Qgp} and a wide range of studies focused on the heavy mesons (for a brief review of theoretical research, see \cite{Mallik,ref2,ref3,ref4,ref5,ref6,ref7,ref8,ref9,ref10,ref11,ref12,Axial_quarkonia,ref13,Yazici}). These particles also have a significant importance in understanding the nature of the spontaneously broken chiral symmetry in dense media.

The theoretical investigations of the physical observables of hadrons at finite temperature will help us in understanding their properties such as decay widths, coupling constants, production rates, decay constants, etc. in LHC and RHIC experiments. Comprehending the decay constants of the heavy mesons has a vital role in the studies of the strong decays of the heavy mesons, as well as in their electromagnetic structures and radiative decay widths \cite{Kumar}. Theoretical predictions of temperature-dependent hadronic quantities also will help us demystify hot and dense QCD matter.

Since hadrons are bound states which are formed beyond perturbative region, one needs to have non-perturbative methods to predict hadronic properties. Because of the increasing amount of energies and luminosities in Pb-Pb collisions, the investigation of the strong interactions of the heavy mesons takes many colleagues' attention \cite{tempp}. There exists a wide literature on the calculations of masses, decay constants or form factors of heavy mesons at vacuum using non-perturbative approaches \cite{re1,re2,re3,re4,re5,re6}. Besides, temperature dependencies of masses and decay constants are studied extensively e.g. using the QCD sum rules. Heavy vectors have also been studied at finite temperature with QCD sum rules, in addition using the maximum entropy method (MEM) to extract the spectral function \cite{phil1,phil2,phil3}.  The AdS/QCD method is applied to the heavy mesons in hot and dense medium as well \cite{Ads1, Ads2}. A remarkable spectral change in charmonium around the critical temperature is shown in these references. However, the thermal behaviours of the hadronic form factors have been investigated rarely \cite{Yazici}. In order to calculate the thermal strong form factors, one must have the meson masses and decay constants as functions of temperature. One of the motivations of this study is mainly to obtain these functions for future studies on thermal form factor calculations of the heavy mesons. 

 In the QCD sum rules approach \cite{11} the properties of heavy mesons are modified mostly through gluon condensates. In the thermal QCD sum rules \cite{12}, it is assumed that the quark-hadron duality and the Operator Product Expansion (OPE) remain valid up to a limited temperature, but the quark and the gluon condensates have new thermal expectation values because of the hadronic medium \cite{13,14,15,16}.
 
 In this paper, the masses and the decay constants of heavy vector and axial vector quarkonia are re-calculated in the framework of the thermal QCD sum rules. These thermal mass and decay constant functions are needed for a future study on the strong form factors of heavy mesons at finite temperature. The originality of this study is the calculation of temperature-dependent masses and decay constants of vector and axial vector $\overline{c}b$ ground states. The observation of these mesons have not been reported yet, but theoretical calculations at $T=0$ such as relativistic \cite{Bc1,Bc2} and non-relativistic quark models \cite{Bc3}, QCD lattice \cite{Bc4,Bc5,Bc6} and QCD sum rules at vacuum \cite{WangBc} have been conducted.

The paper is organized as follows: in the  next section, the thermal QCD sum rules for the masses and the leptonic decay constants of the related mesons are obtained using the two-point correlation functions and the operator product expansion. In Section III, the selected parameters, numerical predictions and the comparisons of the results with the previous studies are given. Last section is devoted to the conclusion.

%%%
%%%
\section{Thermal QCD Sum Rules for the Masses and Decay Constants}
In this section, the temperature-dependent version of the two-point correlation function is evaluated,
\begin{eqnarray}\label{correl.func.1}
\Pi_{\mu\nu}(q,T) =i\int d^{4}x e^{iq \cdot x}{\langle} {\cal T}[J_{\mu} (x) J_{\nu}^{\dagger} (0)]{\rangle},
\end{eqnarray}
where $\cal T$ denotes the time ordering operator on the product of the parameters in the brackets, $T$ is temperature, $x=x_\mu$ and $q=q_\mu$ are the four coordinate and the four momentum vectors, $J_{\mu}(x)$ is the isospin averaged interpolating current of the mesons:

\begin{eqnarray}\label{interpolating1}
\ J_{\mu}^{v}(x)=\overline{Q}(x)\gamma_{\mu}Q(x),
\end{eqnarray}
\begin{eqnarray}\label{interpolating2}
\ J_{\mu}^{a}(x)=\overline{Q}(x)\gamma_{\mu}\gamma_{5}Q(x).
\end{eqnarray}

The currents $J_{\mu}^{v}(x)$ and $J_{\mu}^{a}(x)$ interpolate the heavy vector and axial vector mesons, respectively. The thermal average of any operator $A$ is given as:
\begin{eqnarray}\label{A}
{\langle}A {\rangle}=\frac{Tr(e^{-\beta H} A)}{Tr e^{-\beta H}},
\end{eqnarray}
where $H$ is the QCD Hamiltonian and $\beta=1/T$ is the inverse temperature. If the hadronic matter's reference frame is chosen at rest, Lorentz invariance is broken down. Hence, by using four velocity vector $u_\mu=(1,0,0,0)$, the Lorentz invariant parameters can be redefined. With these parameters, the thermal correlation function is obtained. For the hadronic representation, the correlation function is saturated with the currents having the same quantum numbers of the related mesons. The QCD representation of the correlation function can be written via OPE in terms of e.g. quark masses, quark and gluon condensates which represent the internal structure of the meson and quark-gluon-vacuum interactions. After equating these two different pictures of the correlation function, the QCD sum rules are obtained. However, the higher states' contribution should be suppressed and the Borel transformation is the main concept for this suppression. As a result, for the hadronic representation:
 
\begin{eqnarray}\label{phen1}
\Pi_{\mu\nu}^{hadron}(q,0)=\frac{\langle 0\mid J_{\mu}(0) \mid M(p,\lambda)\rangle \langle M(p,\lambda)\mid J_{\nu}(0) \mid 0\rangle} {m^{2}_{M}-q^2 }
&+& \cdots ,
\end{eqnarray}

where ($\cdots$) represents the excited states and the continuum, $m_M$ is the mass of the heavy meson. The matrix element which creates the meson current from the vacuum is defined in terms of decay constants and masses as:

\begin{eqnarray}\label{cur}
\langle 0\mid J_{\mu}(0) \mid M(p,\lambda)\rangle = f_{M} m_{M} \epsilon_{\mu}^{(\lambda)},
\end{eqnarray}

where $\epsilon_{\mu}^{(\lambda)}$ is the polarization vector. Summation rule over polarization vectors is given as:

\begin{eqnarray}\label{polar}
\sum_{\lambda}\epsilon_{\mu}^{(\lambda)\star}\epsilon_{\nu}^{(\lambda)}= (-g^{\mu\nu}+\frac{q_\mu q_\nu}{m_M^2}).
\end{eqnarray}

Note that Eqs. 5-7 are also valid at finite temperature. After some algebraic operations, the thermal hadronic representation is obtained as:

\begin{eqnarray}\label{phen1}
\Pi_{\mu\nu}^{hadron}(q,T)= \frac{f_M^2(T) m_M^2(T)}{m_M^2(T)-q^2}(-g^{\mu\nu}+\frac{q_\mu q_\nu}{m_M^2(T)})+\cdots,
\end{eqnarray}
where $f_M(T)$ and $m_M(T)$ are temperature-dependent decay constants and masses of the mesons, respectively.

 The QCD representation of the correlation function is obtained in deep Euclidean region, $q^2\ll-\Lambda^{2}_{QCD}$, by using OPE where the high energy and low energy regions are separated:

\begin{eqnarray}\label{correl.func.QCD1}
\Pi_{\mu\nu}^{QCD}(q^2,T) =\Pi_{\mu\nu}^{pert}(q^2,T)+\Pi_{\mu\nu}^{nonpert}(q^2,T).
\end{eqnarray}

\begin{figure}[h!]
\begin{center}
\includegraphics[width=5cm]{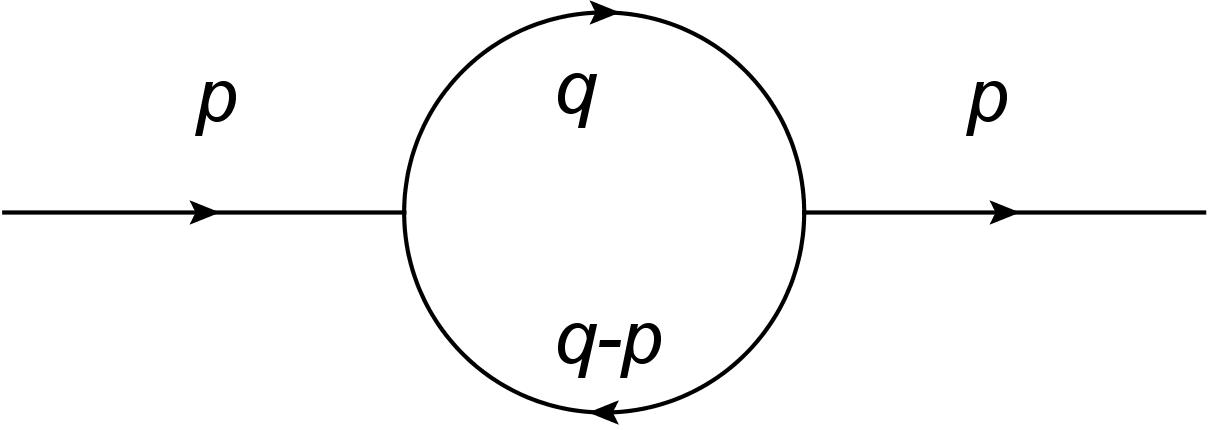}
\end{center}
\caption{Bare loop diagram for the perturbative contribution.}
\label{bare}
\end{figure}

\begin{figure}[h!]
\begin{center}
\includegraphics[width=5cm]{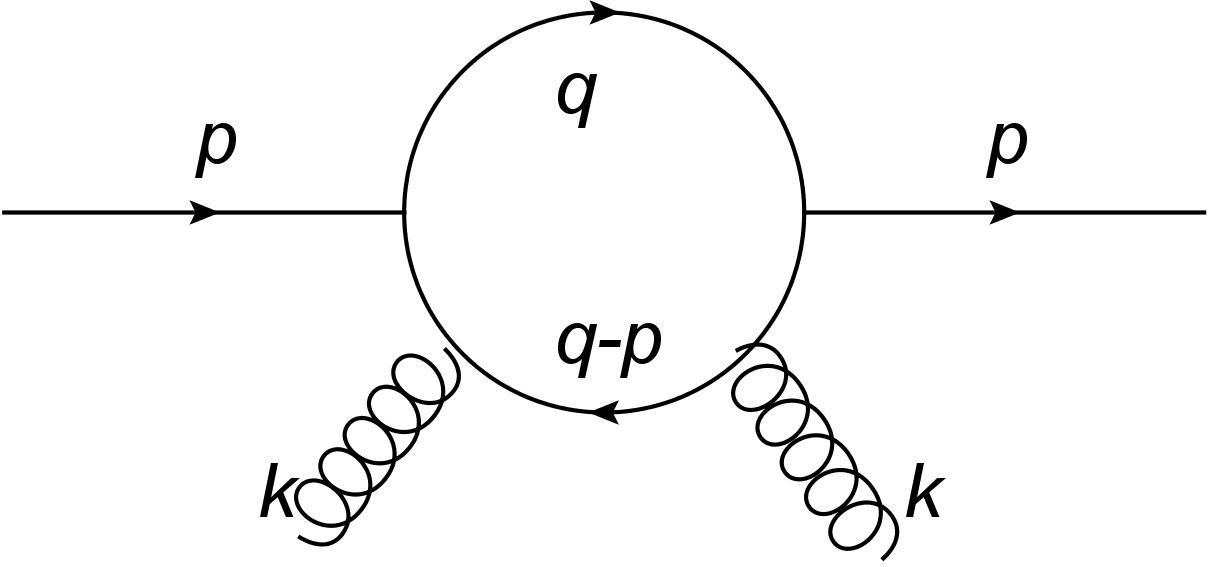}
\includegraphics[width=5cm]{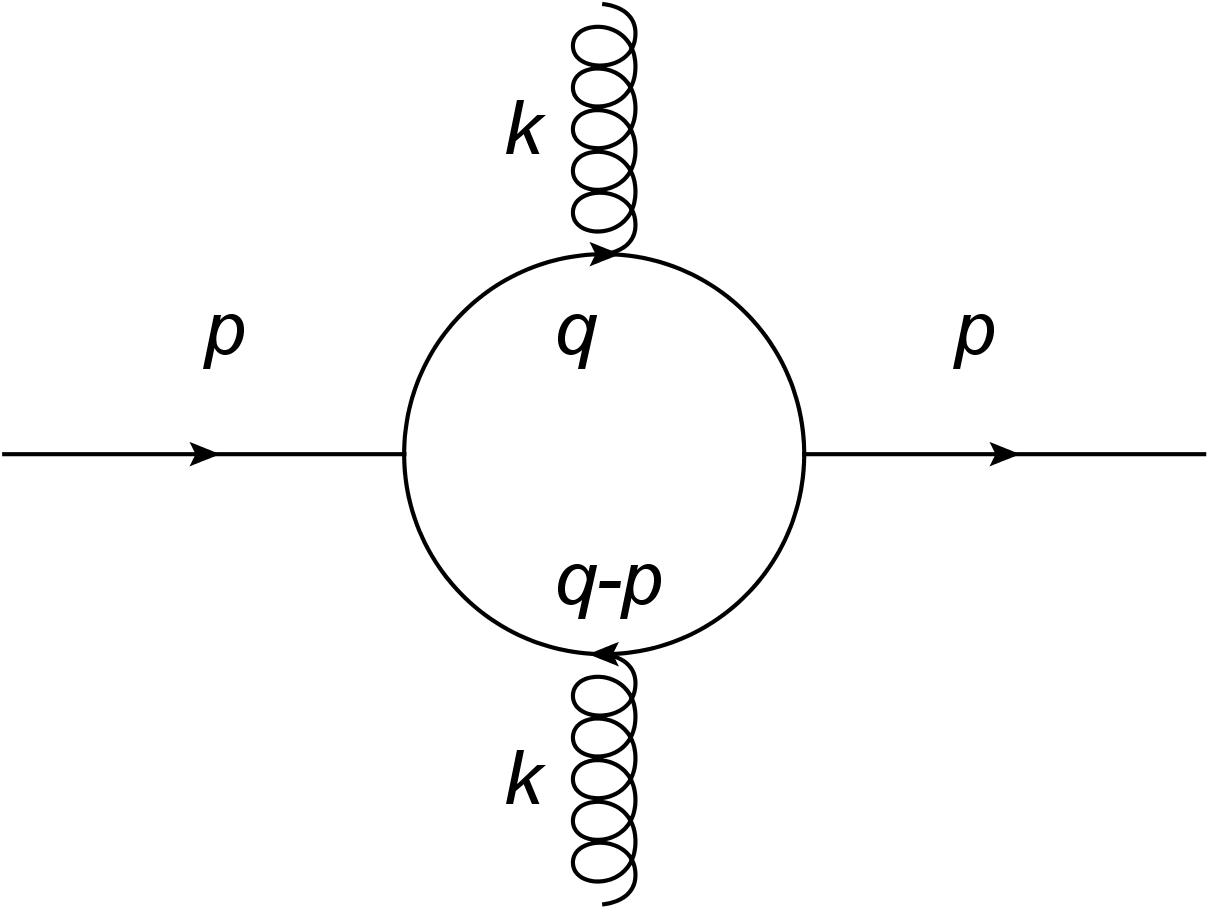}
\includegraphics[width=5cm]{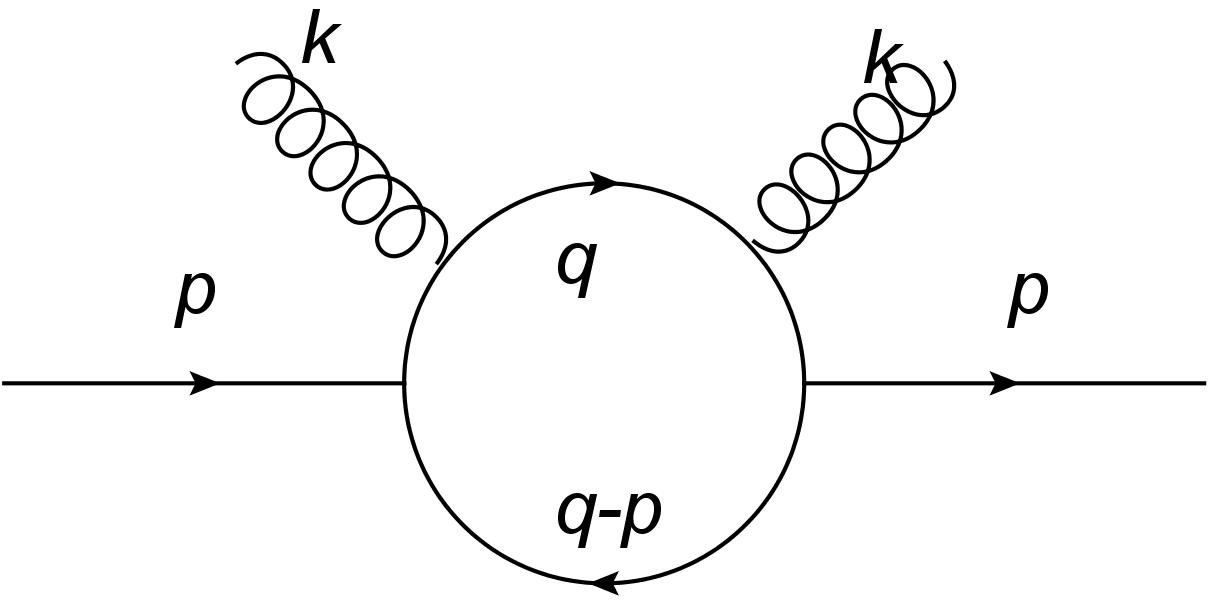}
\end{center}
\caption{Gluon condensate diagrams for the non-perturbative contributions.}
\label{np}
\end{figure}

The perturbative part is represented by the bare loop diagram (Fig \ref{bare}) and is calculated in terms of a dispersion integral: 

\begin{eqnarray}\label{pert.QCD1}
\Pi_{\mu\nu}^{QCD}(q^2,T) =\int \frac{ds\rho^{\mu\nu}(s,T)}{s-q^2}+\Pi_{\mu\nu}^{nonpert}(q^2,T),
\end{eqnarray}
where $\rho^{\mu\nu}$ is the thermal spectral density. By using Cutkosky's rule and after some straightforward algebra, the thermal spectral density is obtained as:

\begin{eqnarray}\label{rhoper}
\rho(s,q^2,T)&=& \frac{N_c}{2\pi^2} \sqrt{\frac{m_1^2+s-m_2^2}{4s}-m_2^2}\left[-\frac{4(m_2^2-m_1^2+s)}{s•}+8\left(-\frac{m_2^2}{3s}+\frac{(m_2^2-m_1^2+s)^2}{3s^2}\right)\right] ,
\end{eqnarray}

where $N_c=3$ is the number of colors, $m_1$ and $m_2$ are the valence quark masses of the related heavy meson. Here, $p_\mu p_\nu$ is chosen as the structure and the terms which are proportional to $p_\mu p_\nu$ are written.

For the non-perturbative part, the quark condensate contributions are suppressed by the inverse powers of the quark masses. For the heavy-heavy mesons, the quark condensate contributions are extremely weak and can be neglected. Hence, the gluon condensates play the major role in the calculations and the non-perturbative contribution comes mainly from the two-gluon condensates which are shown in Fig \ref{np}. Fock-Schwinger gauge $x^{\mu}A^{a}_{\mu}(x)=0$ is considered to calculate the gluon condensates. In the calculations, the following expressions are used for the vacuum gluon field and the quark-gluon-quark vertices:
\begin{eqnarray}\label{Amu}
A^{a}_{\mu}(k)=-\frac{i}{2}(2 \pi)^4 G^{a}_{\rho
\mu}(0)\frac{\partial} {\partial k_{\rho}}\delta^{(4)}(k),
\end{eqnarray}
\begin{eqnarray}\label{qgqver}
\Gamma=ig \frac{\lambda^{a}}{2}\gamma_{\mu}A^{a}_{\mu}(k),
\end{eqnarray}
where $k$ is the four momentum of the gluons.

Considering the Lorentz covariance at finite temperature, the expectation value ${\langle}Tr^{c}G_{\alpha \beta}G_{\mu \nu}{\rangle}$ must be known and it is given as \cite{15}:

\begin{eqnarray}\label{qgqver}
{\langle} Tr^{c}G_{\alpha \beta}G_{\mu \nu} {\rangle}&=&\frac{1}{24}\left(g_{\alpha\mu}g_{\beta\nu}-g_{\alpha\nu}g_{\beta\mu}\right){\langle} G^{a}_{\lambda \sigma}G^{a\lambda\sigma} {\rangle}\nonumber\\
&+&\frac{1}{6}\left[g_{\alpha\mu}g_{\beta\nu}-g_{\alpha\nu}g_{\beta\mu}-2(u_{\alpha}u_{\mu}g_{\beta\nu}-u_{\alpha}u_{\nu}g_{\beta\mu}-u_{\beta}u_{\mu}g_{\alpha\nu}+u_{\beta}u_{\nu}g_{\alpha\mu})\right]{\langle} u^{\lambda}\Theta_{\lambda\sigma}^{g}u^{\sigma} {\rangle}.
\end{eqnarray}

After the Borel transformation, the non-perturbative contribution is obtained as:
\begin{eqnarray}\label{qgqver}
\hat{B}\Pi^{QCD}(q^2,T)&=&\int_{0}^{1} dx \frac{1}{12M^2\pi^2(x-1)^3x^3}\exp[\frac{\frac{m_2^2}{x-1}-\frac{m_2^2}{x}}{M^2}]\nonumber\\
&& \times \Big[3 A(T) \Big\{ 2m_1^4 (x-1)^5(x+1)+m_1^2(x-1)^2x \Big[ m_2^2x(-4x^2+3x+2)+2M^2(x-1)^2(4x^2-3) \Big]
\nonumber\\
&+& x^3\Big[m_2^4x^2(2x-3)+6M^2(x-1)^3(2x^2-2x-1)+m_2^2M^2x(-8x^3+23x^2-19x+4)\Big] \Big\}
\nonumber\\
&+& B(T) \Big\{ -2m_1^4(x-1)^4(6x^2-4x-1)+x^2\Big[m_2^4x^2(-12x^2+17x-4)
\nonumber\\
&-& M^2(x-1)^2(9x^4-5x^3-36x^2+28x+8)+m_2^2M^2x(27x^4-92x^3+123x^2-80x+22)\Big]
\nonumber\\
&-& m_1^2(x-1)^2x\Big[m_2^2x(-24x^2+25x-2)+M^2(27x^4-59x^3+44x^2-8x-4)\Big] \Big\} \Big] ,
\end{eqnarray}
where $M^2$ is the Borel parameter, $A(T)$ and $B(T)$ are parameters related to traceless gluonic part of the energy density $\Theta_{\lambda\sigma}^g$: 

\begin{eqnarray}\label{AtBt}
A(T)=\frac{1}{24}{\langle} G^{a}_{\lambda \sigma}G^{a\lambda\sigma} {\rangle} +\frac{1}{6}{\langle} u^{\lambda}\Theta_{\lambda\sigma}^{g}u^{\sigma} {\rangle}
\nonumber\\
B(T)=\frac{1}{3}{\langle} u^{\lambda} \Theta_{\lambda\sigma}^g u^{\sigma} {\rangle} ,
\end{eqnarray}

After equating the hadronic and the QCD sides of the correlation functions, the QCD sum rules for the heavy-heavy vector and axial vector mesons' masses and decay constants are obtained. Using the Borel transformation to suppress the excited states contributions as well as continuum subtraction, we get:
\begin{eqnarray}\label{aasum}
f_M^2(T)m_M^2(T)\exp\Big[-\frac{m_M^2(T)}{M^2}\Big]= \int_{(m_1+m_2)^2}^{s_0(T)}ds \rho(s)\exp\Big[-\frac{s}{M^2}\Big]+\hat{B}\Pi_{\mu\nu}^{QCD},
\end{eqnarray}
where $s_0(T)$ is the thermal continuum threshold. In order to obtain the temperature-dependent mass expression, we apply derivative with respect to $-1/M^2$ to the both sides of Eq. \ref{aasum} and divide by itself:
\begin{eqnarray}\label{aasum2}
m_M^2(T)= \frac{\int_{(m_1+m_2)^2}^{s_0(T)}ds \rho(s)\exp[-\frac{s}{M^2}]+\frac{-d}{d(1/M^2)}\Pi^{QCD}}{\int_{(m_1+m_2)^2}^{s_0(T)}ds \rho(s)\exp[-\frac{s}{M^2}]+\hat{B}\Pi_{\mu\nu}^{QCD}} .
\end{eqnarray}

The continuum threshold expression is given as \cite{Cheng,Miller}:

\begin{eqnarray}\label{continuum1}
s_{0}(T)=s_{0}\left[ 1- \left( \frac{T}{T_{c}}\right)^{8} \right]+(m_{1}+m_{2})^2\left( \frac{T}{T_{c}}\right)^{8}~ ,
\end{eqnarray}

where the critical temperature value is considered as $T_c=0.197GeV$.

\section{Numerical Results}

In this section we present the numerical results of  the sum rules for the  masses and the decay constants of the heavy-heavy mesons based on temperature-dependent relations. For this aim, the following input parameters are used,
$m_c=(1.275\pm0.05)GeV$, $m_b=(4.66\pm0.1)GeV$ \cite{pdg} and
${\langle}0\mid\frac{1}{\pi}\alpha_s G^2  \mid 0 {\rangle}=0.012GeV^4$
\cite{11}. For the gluonic part of the energy density, in the rest frame of hot medium, the lattice results are parametrized as \cite{Cheng,Miller,20}:
\begin{eqnarray}\label{tetag}
\langle \Theta^{g}_{00}\rangle=T^4~ exp\left[113.87T^2-12.2T\right]-10.14T^5(GeV^4).
\end{eqnarray}

This expression is valid for $T>130MeV$. The temperature-dependent gluon condensate is used in \cite{Ayala} as:
\begin{eqnarray}\label{G2TLattice}
\langle G^2\rangle=\langle
0|G^2|0\rangle\left[1-1.65\left(\frac{T}{T_{c}}\right)^{8.735}+0.05\left(\frac{T}{T_{c}}\right)^{0.72}\right].
\end{eqnarray}
where $\langle0|G^2|0\rangle$ stands for the gluon condensate expectation value for the vacuum. For the working regions of the auxiliary parameters which are the Borel mass $M^2$ and the continuum threshold $s_0$, it is expected that the observable quantities to be independent of these parameters. While determining the Borel mass working region, the higher states and the continuum contributions should be suppressed. The continuum threshold is generally associated with the energy of the first excited state of the related meson (for a brief discussion see \cite{ref13}). Under these conditions, working regions of $M^2$ and $s_0$ are given in Table \ref{tab:borel}. While determining the continuum thresholds, the predictions for the mass difference of vector and axial vector mesons having the same quark contents, based on the non-relativistic renormalization group are also considered:  $m_{AV}-m_V=(50\pm17)MeV$ \cite{Penin}.

\begin{table}[h]
\renewcommand{\arraystretch}{1.5}
\addtolength{\arraycolsep}{3pt}
$$
\begin{array}{|c|c|c|c|c|c|c|}
\hline 
  ~ & \Upsilon &  B_{c}^{1^-} & J/\psi & \chi_{b1} & B_{c}^{1^+} & \chi_{c1} \\
\hline
  M^2 (GeV^2) &  10-20   &  6-10   &  6-10  & 14-20 & 10-14 & 15-25 \\
 \hline
 s_0 (GeV^2)  &  102\pm2  &  45\pm1  &  11\pm1  &  110\pm5 & 52\pm1 & 16\pm1 \\
  \hline 
  \hline
\end{array}
$$
\caption{The Borel mass parameters and the continuum threshold parameters used in the calculations.} \label{tab:borel}
\renewcommand{\arraystretch}{1}
\addtolength{\arraycolsep}{-1.0pt}
\end{table}

\begin{figure}[h!]
\begin{center}
\includegraphics[width=8cm]{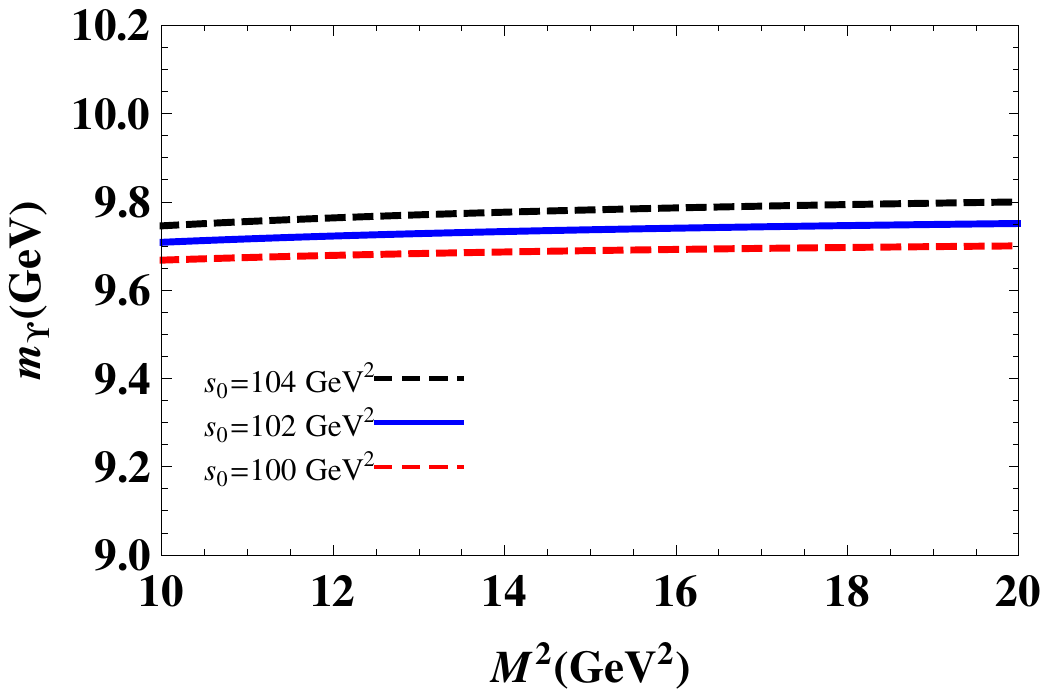}
\includegraphics[width=8cm]{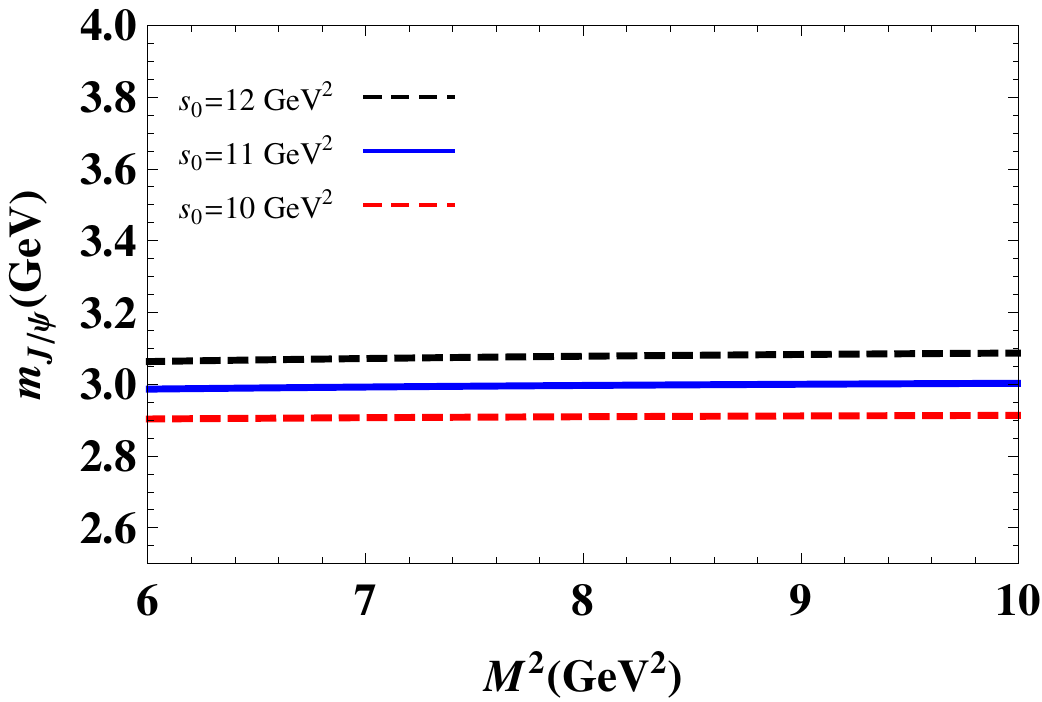}
\includegraphics[width=8cm]{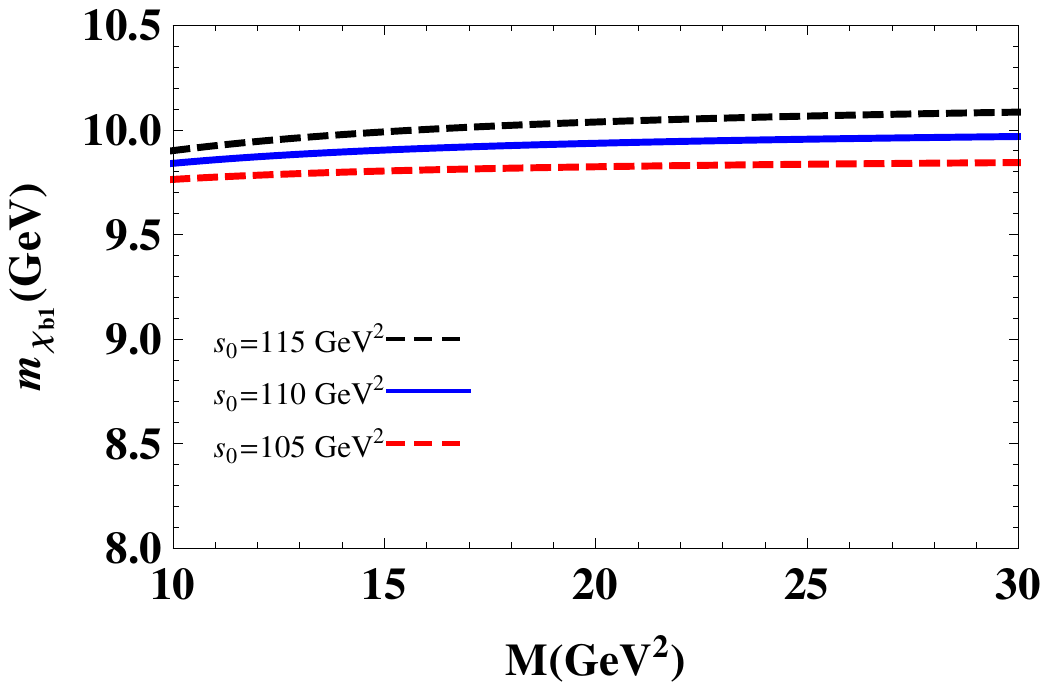}
\includegraphics[width=8cm]{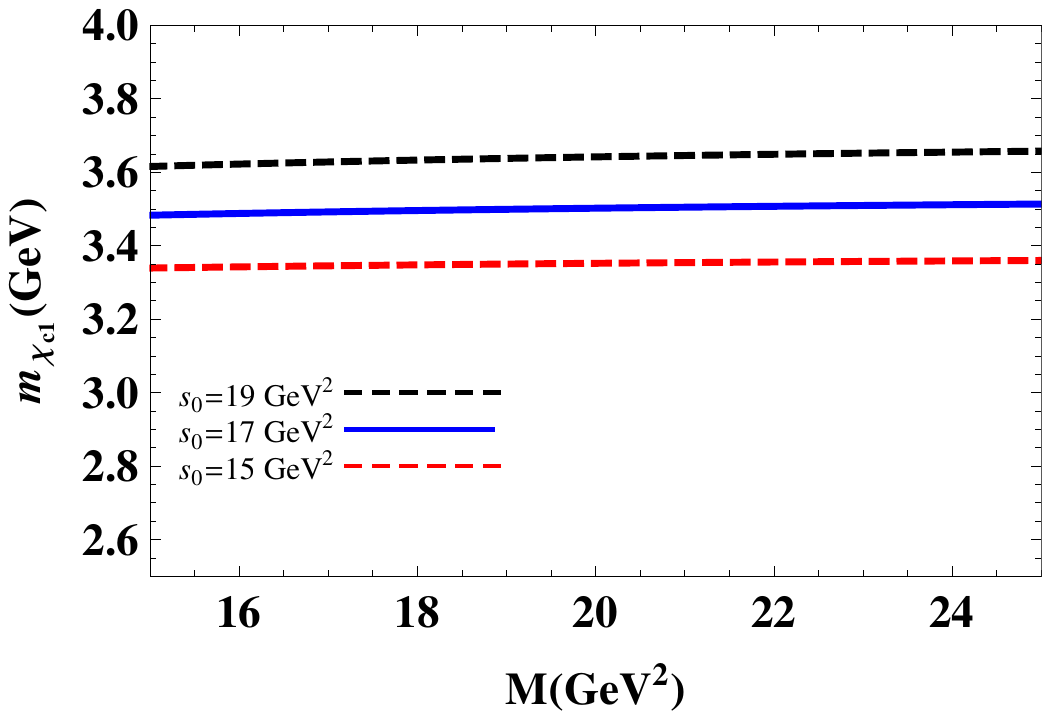}
\end{center}
\caption{The dependencies of vector and axial vector quarkonia masses on $M^2$.}
\label{fig1}
\end{figure}

\begin{figure}[h!]
\begin{center}
\includegraphics[width=8cm]{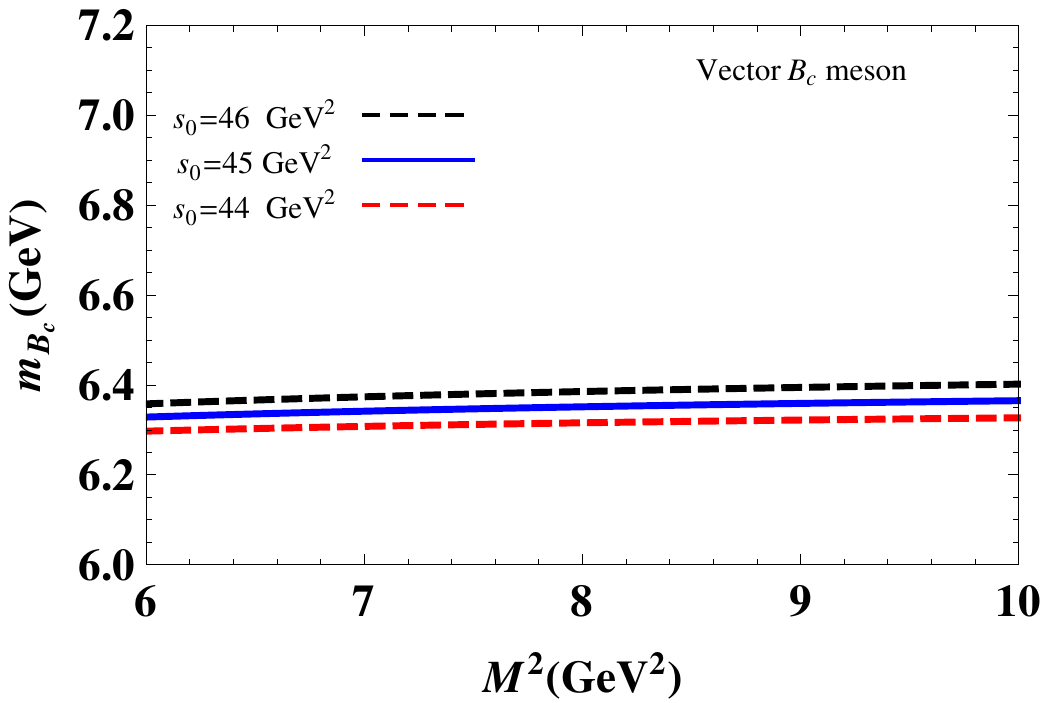}
\includegraphics[width=8cm]{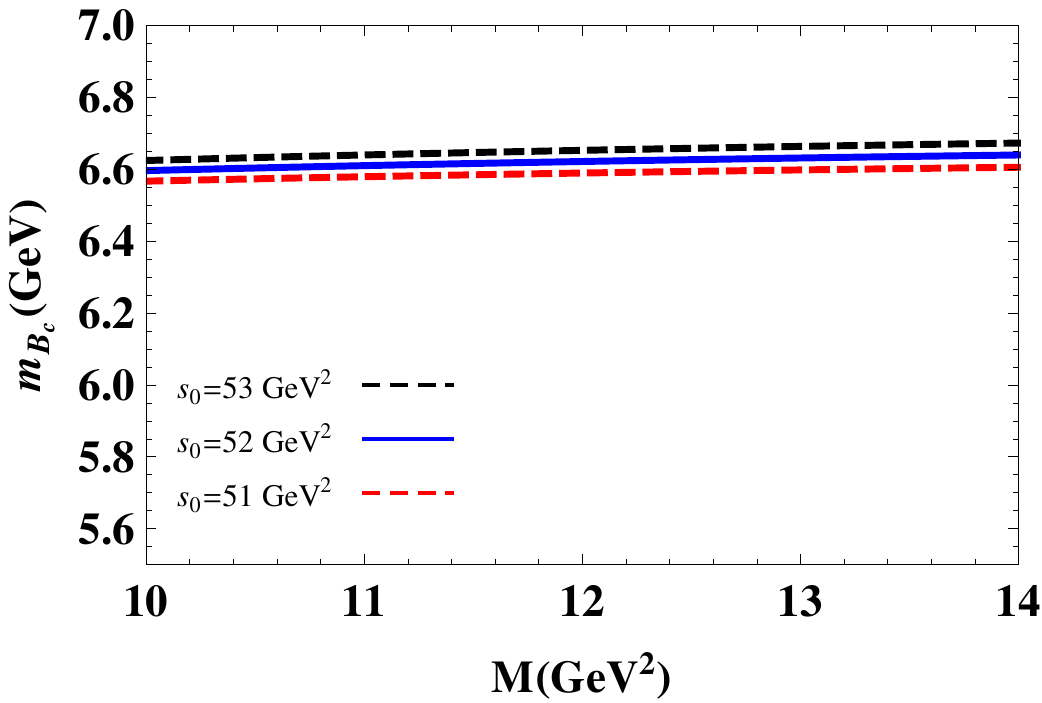}
\end{center}
\caption{The dependencies of vector and axial vector $B_c$ masses on $M^2$.}
\label{fig2}
\end{figure}

In Figs \ref{fig1}-\ref{fig2} the Borel mass parameters for the heavy quarkonia and $B_c$ mesons are presented, respectively. It is seen from these figures that the masses have good stabilities with respect to selected $M^2$ regions.

For the zero temperature mass predictions, the comparison of the results of this study and the other theoretical predictions and experimental data are presented in Table \ref{tab:mass}. It is seen that the thermal QCD sum rules results at $T=0$ are in a good agreement with the previous predictions as well as with the experimental data. The predictions of this study for the masses of vector and axial vector $\bar{c}b$ mesons: $m_{B_c^{1-}}=6.35GeV$ and $m_{B_c^{1+}}=6.62GeV$ can be verified in future observations. It is newsworthy that in the previous thermal QCD sum rules studies \cite{ref12,Axial_quarkonia}, the parametrizations of temperature-dependent continuum thresholds and temperature-dependent gluon condensates are different from this study. Still, the results are in good agreement. The accuracy of the masses in this study is calculated by taking into account the specified $s_0$ and $M^2$ regions in Table \ref{tab:borel} and the accuracies of the quark masses.  

\begin{table}[h]
\renewcommand{\arraystretch}{1.5}
\addtolength{\arraycolsep}{3pt}
$$
\begin{array}{|c|c|c|c|c|c|c|}
\hline 
  ~ & m_{\Upsilon}(GeV) &  m_{B_{c}^{1^-}}(GeV) & m_{J/\psi}(GeV) & m_{\chi_{b1}}(GeV) & m_{B_{c}^{1^+}}(GeV)& m_{\chi_{c1}}(GeV) \\
\hline
  Present\ work &  9.74\pm0.06   &  6.35\pm0.05   &  3.0\pm0.1  & 9.92\pm0.13 & 6.62\pm0.06 & 3.5\pm0.1 \\
 \hline
 Experimental\ \hbox{\cite{pdg}} &  9.4603  &  -  &  3.097  &  9.89 & - & 3.51 \\
  \hline 
 Thermal\ QCDSR\ \hbox{\cite{ref12,Axial_quarkonia}} & 9.68\pm0.25 & - & 3.05\pm0.08 & 9.96\pm0.26 & - & 3.52\pm0.11 \\
\hline 
  Thermal\ QCDSR\ (MEM)\ \hbox{\cite{phil1,phil2,phil3}} & 9.56 & - & 3.06 & 10.42 & - & - \\
 \hline
 QCDSR\ at\ vacuum\ \hbox{\cite{WangBc}} & - & 6.337\pm0.052 & - & - & 6.73\pm0.061 & - \\
 \hline
 Lattice\ QCD\ \hbox{\cite{Bc5}} & - & 6.321 & - & - & 6.743 & - \\
  \hline
\end{array}
$$
\caption{The masses of the heavy mesons at $T=0$ from different theoretical approaches and experimental data.} \label{tab:mass}
\renewcommand{\arraystretch}{1}
\addtolength{\arraycolsep}{-1.0pt}
\end{table}

\begin{figure}[h!]
\begin{center}
\includegraphics[width=8cm]{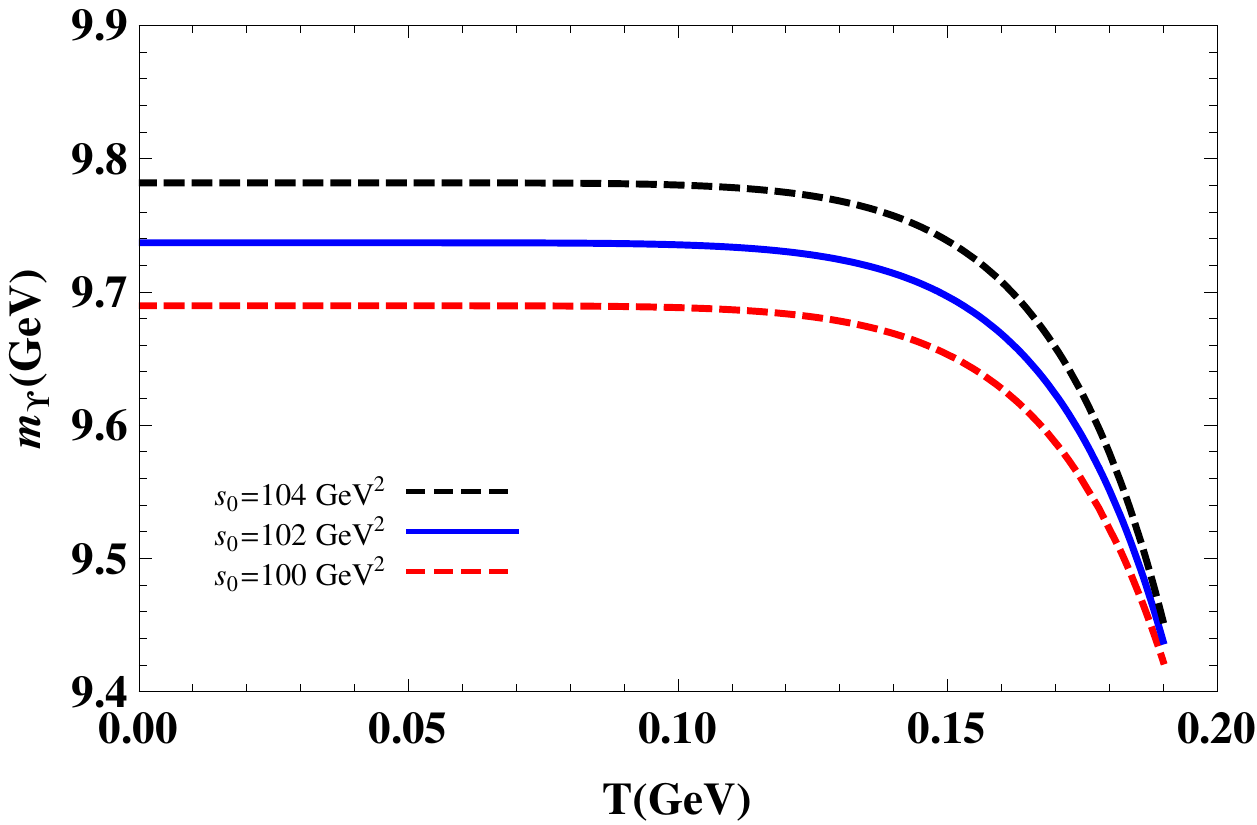}
\includegraphics[width=8cm]{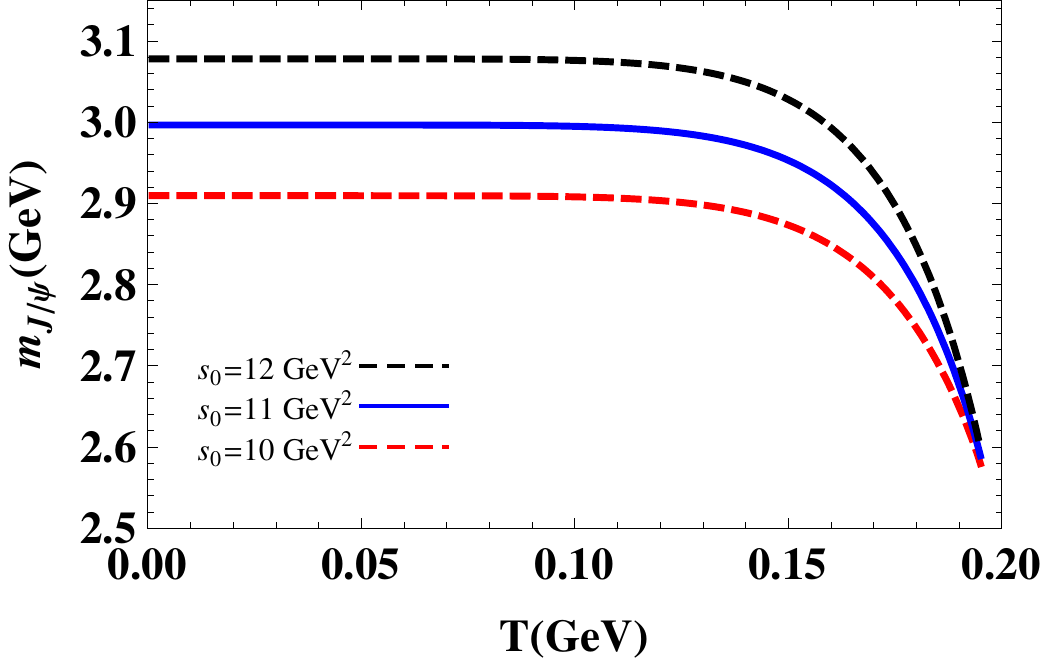}
\includegraphics[width=8cm]{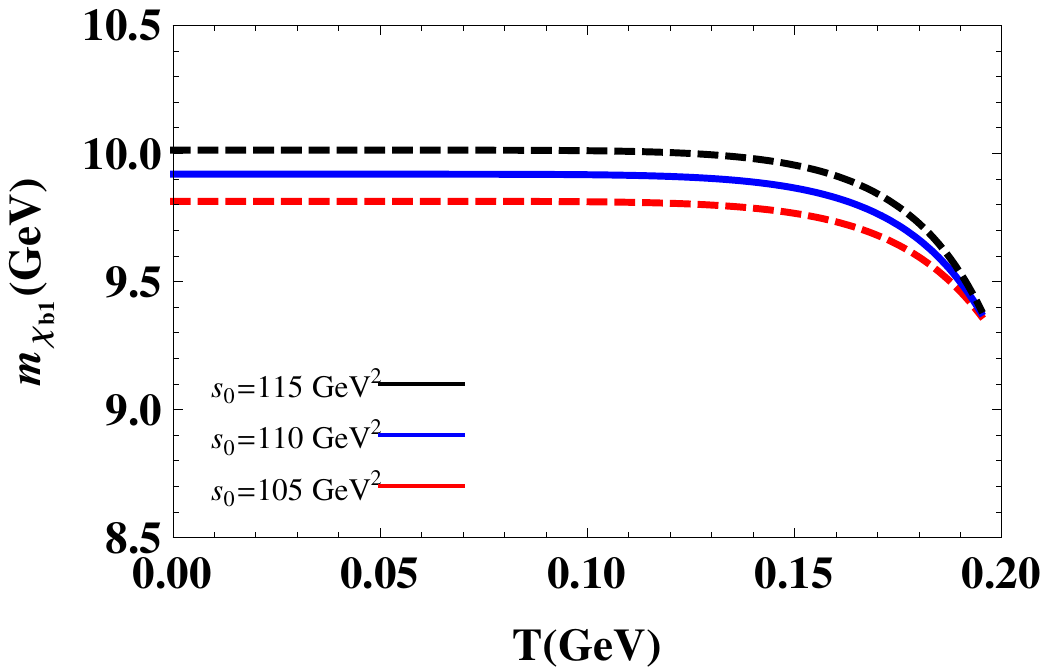}
\includegraphics[width=8cm]{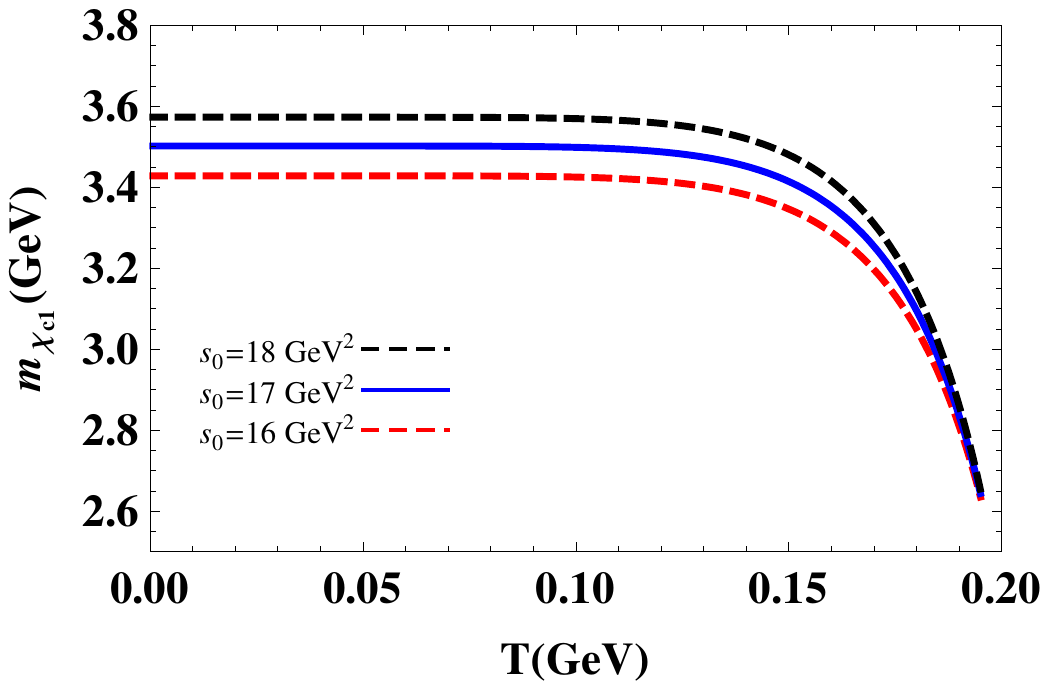}
\end{center}
\caption{Temperature dependencies of heavy-heavy quarkonia masses.}
\label{fig3}
\end{figure}

\begin{figure}[h!]
\begin{center}
\includegraphics[width=8cm]{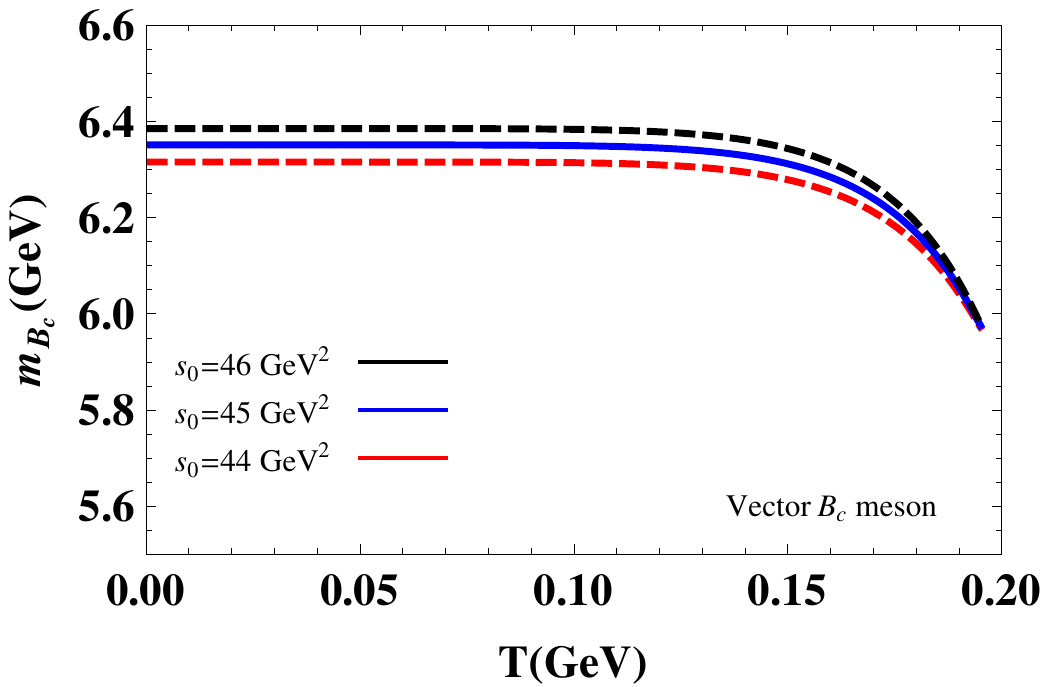}
\includegraphics[width=8cm]{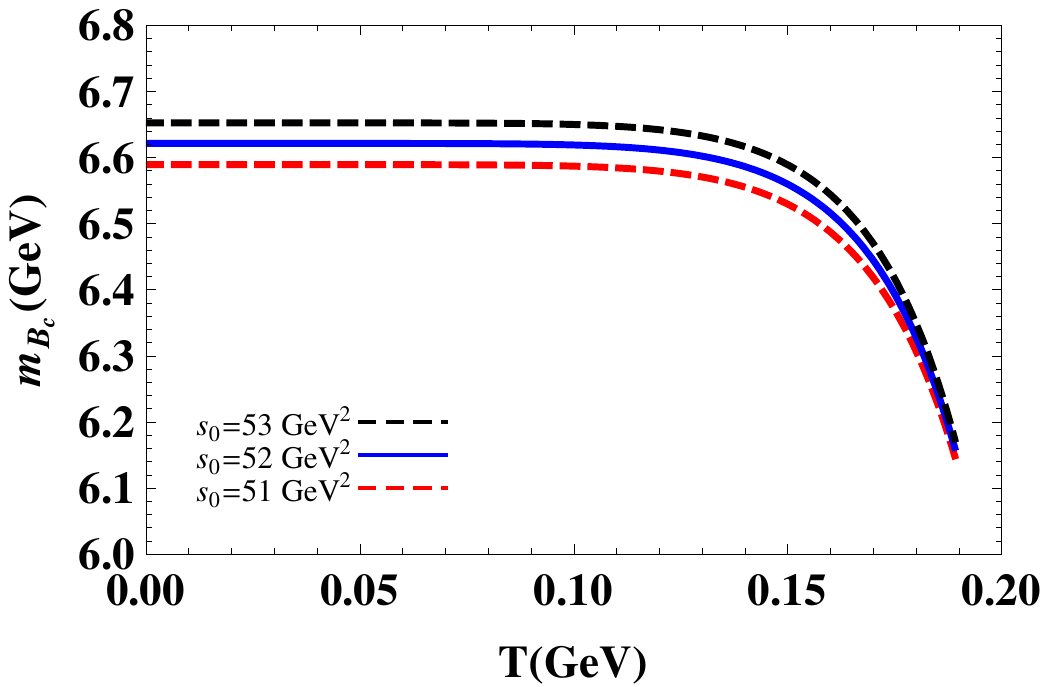}
\end{center}
\caption{Temperature dependencies of vector and axial vector $B_c$ masses.}
\label{fig4}
\end{figure}

The temperature dependencies of the masses are presented in Figs. \ref{fig3} and \ref{fig4}. From these figures, it is seen that the mass values do not change up to $T \simeq 150MeV$. After this point, they start to diminish and around the critical temperature, we see that the masses fall roughly to $~80-85\%$ of their value at vacuum.

\begin{figure}[h!]
\begin{center}
\includegraphics[width=8cm]{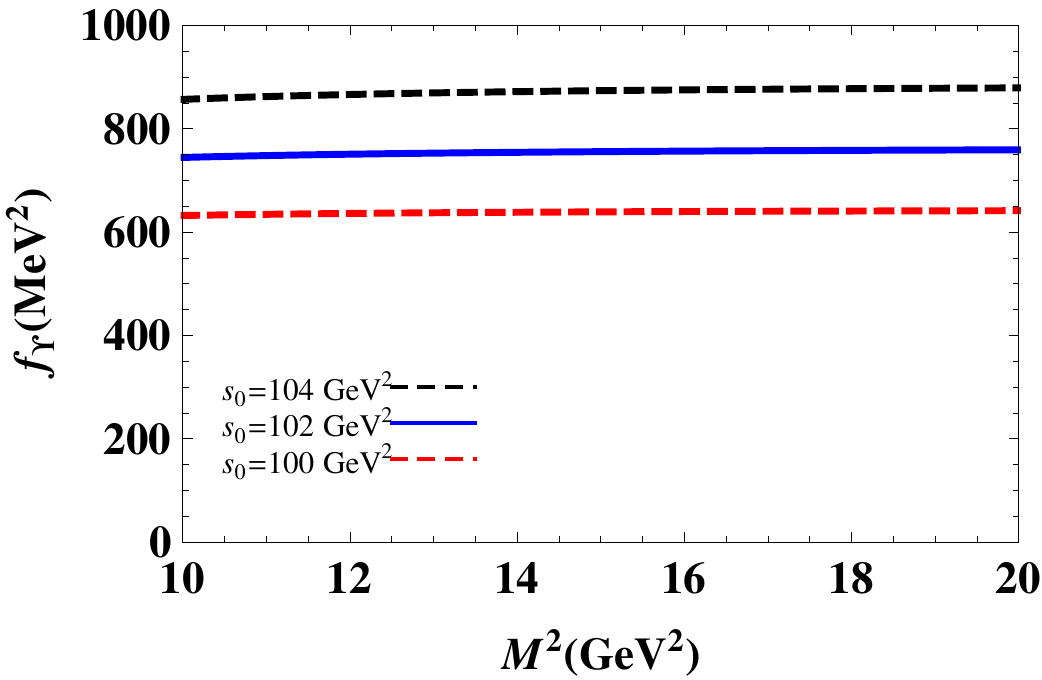}
\includegraphics[width=8cm]{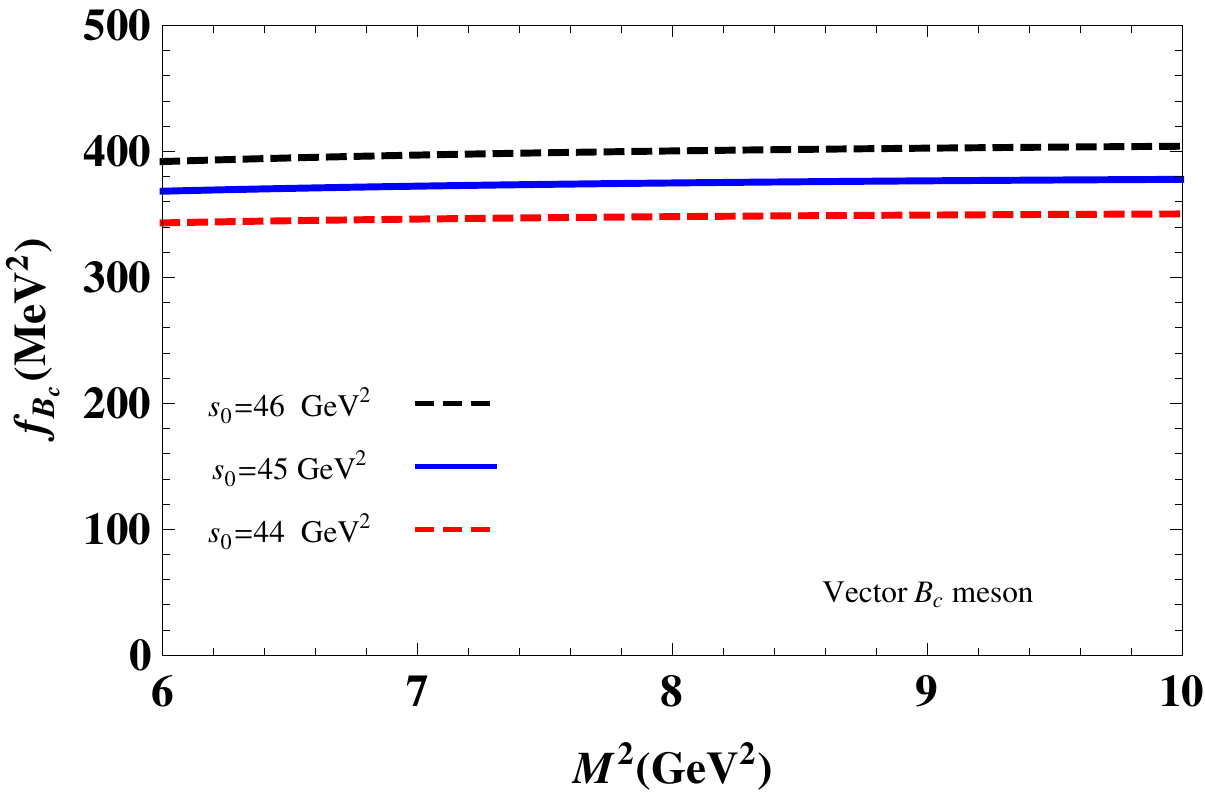}
\includegraphics[width=8cm]{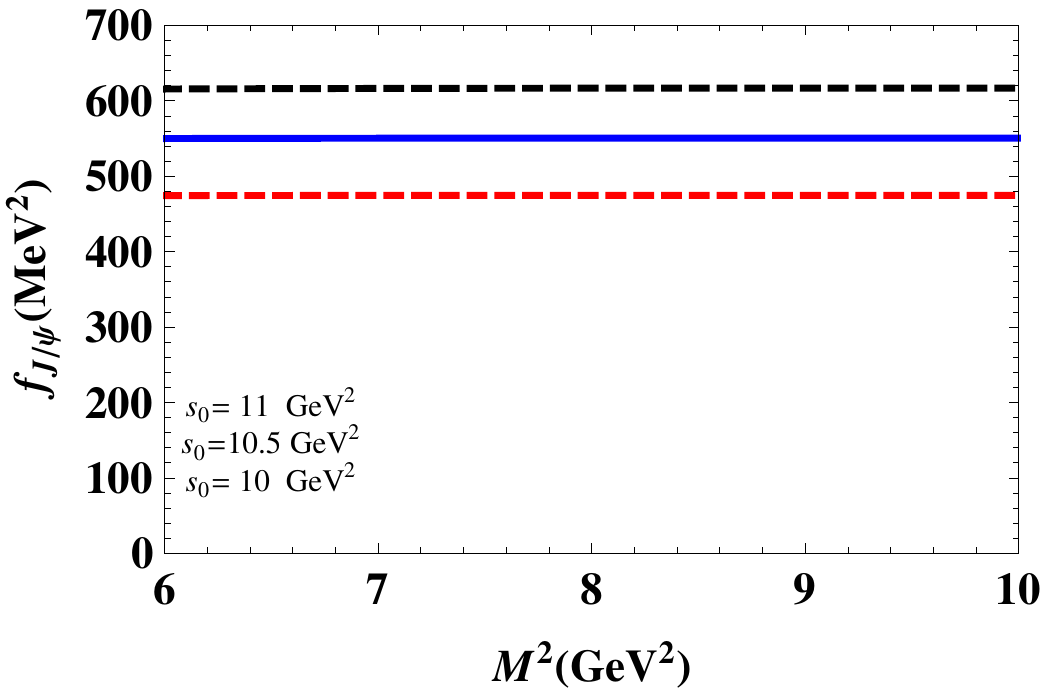}
\includegraphics[width=8cm]{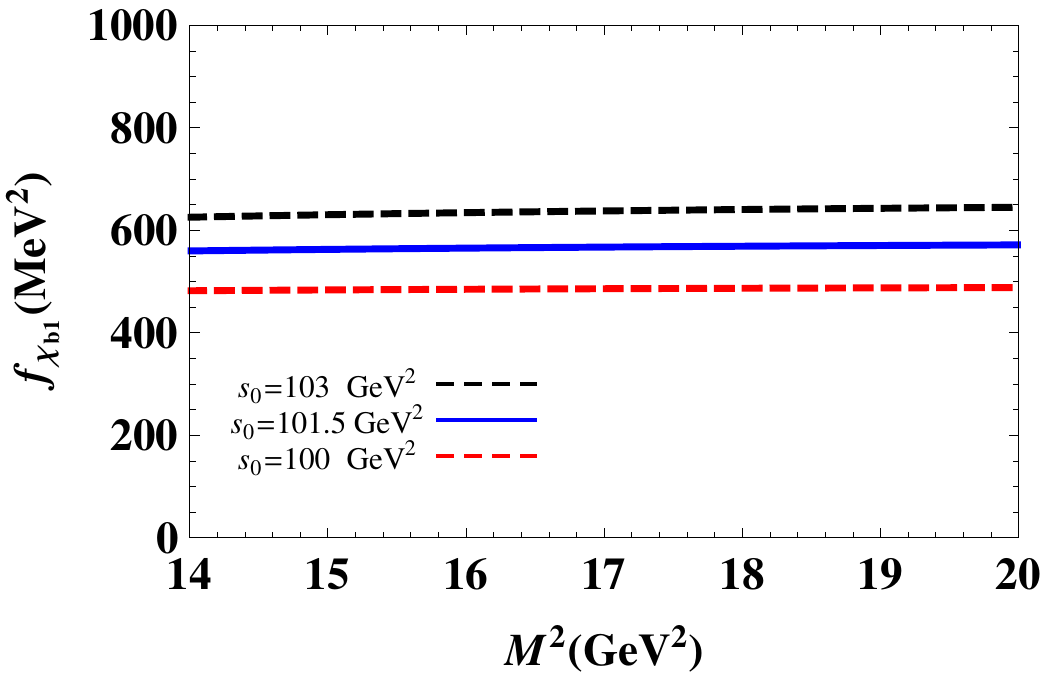}
\includegraphics[width=8cm]{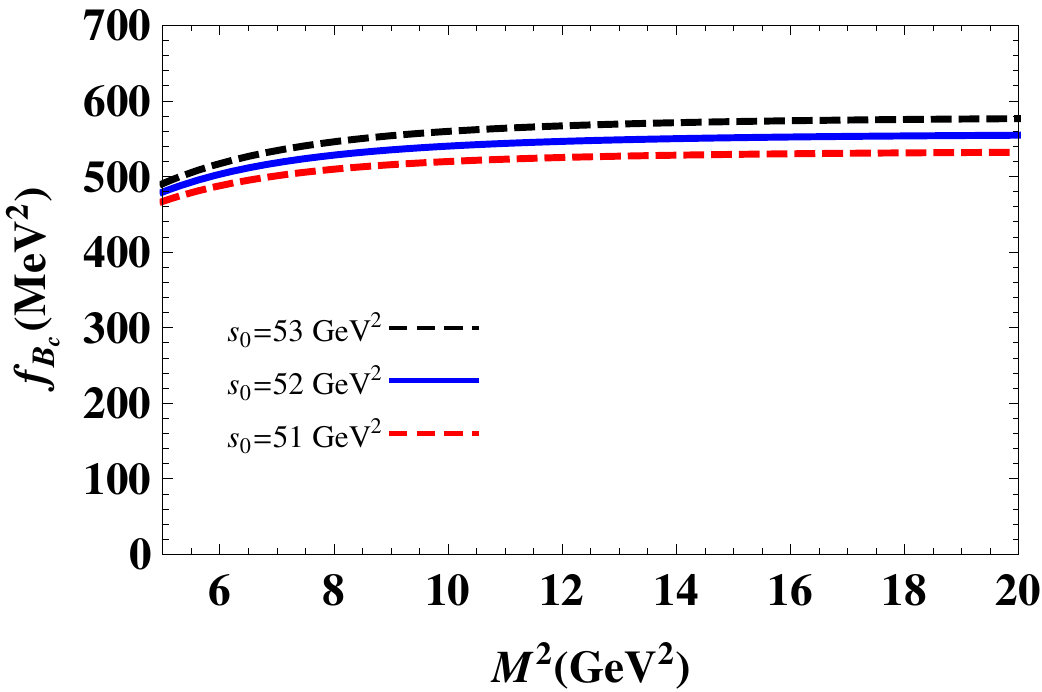}
\includegraphics[width=8cm]{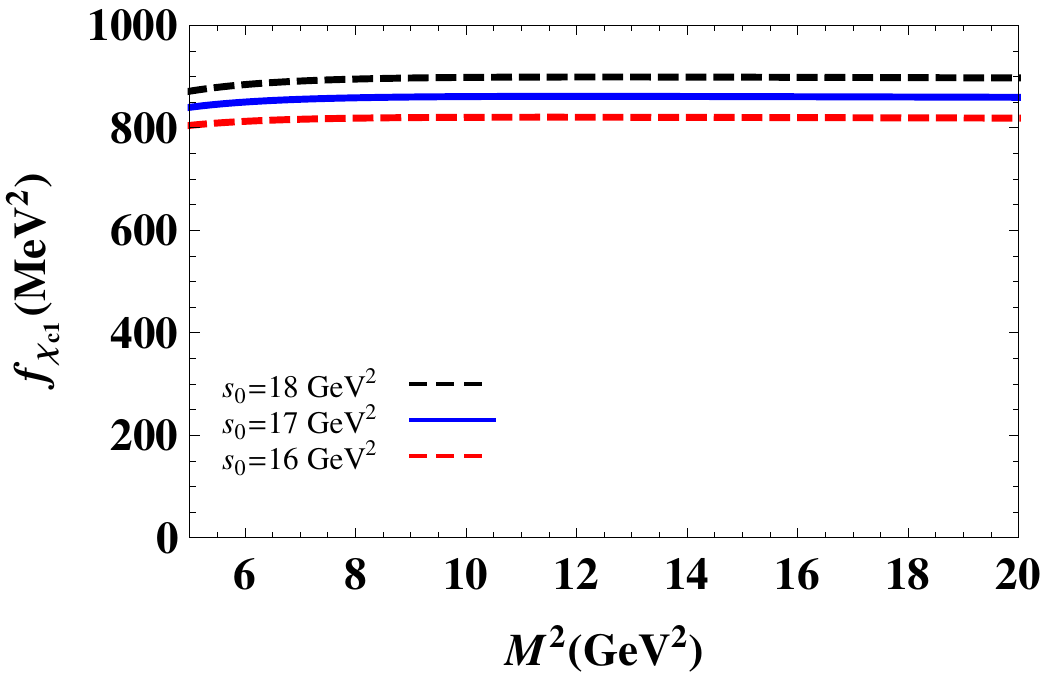}
\end{center}
\caption{The dependencies of vector and axial vector heavy meson decay constants on $M^2$.}
\label{fig5}
\end{figure}

\begin{figure}[h!]
\begin{center}
\includegraphics[width=8cm]{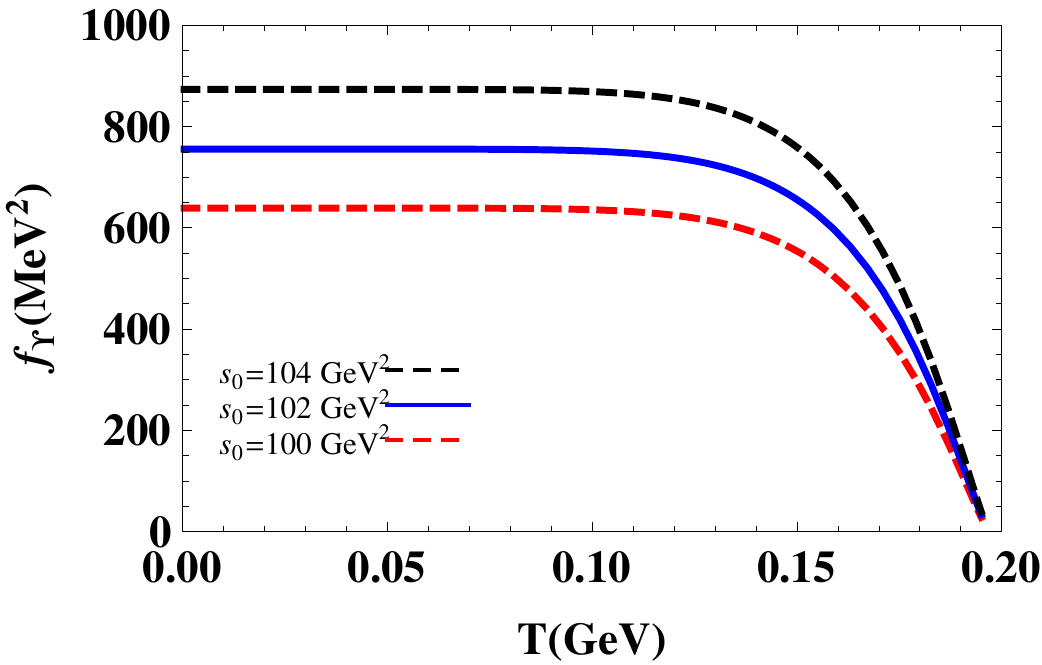}
\includegraphics[width=8cm]{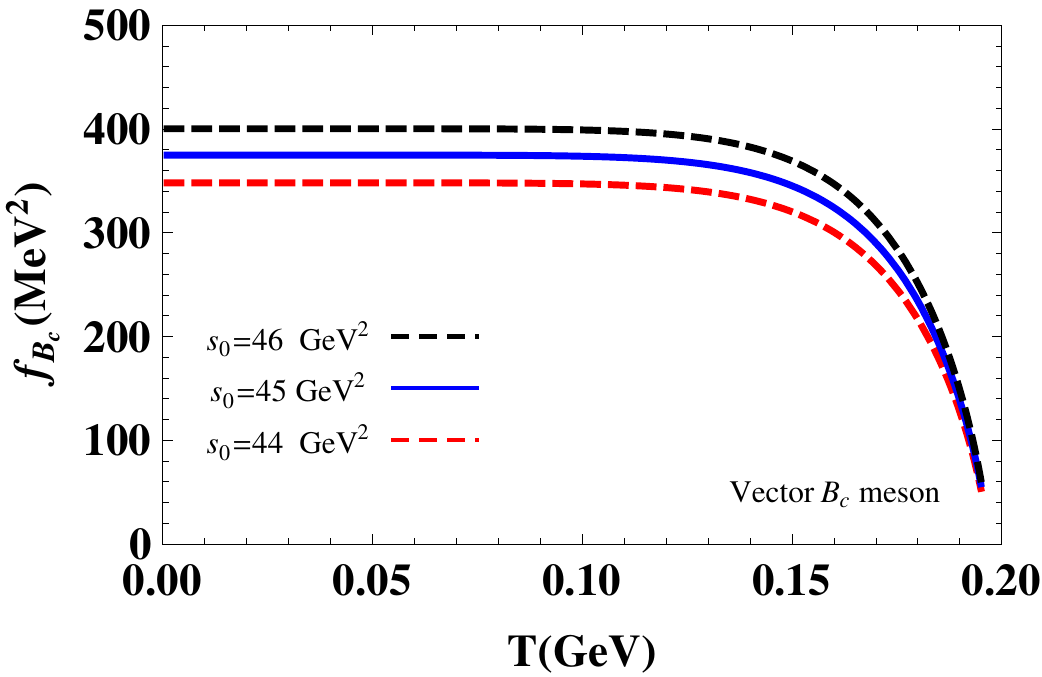}
\includegraphics[width=8cm]{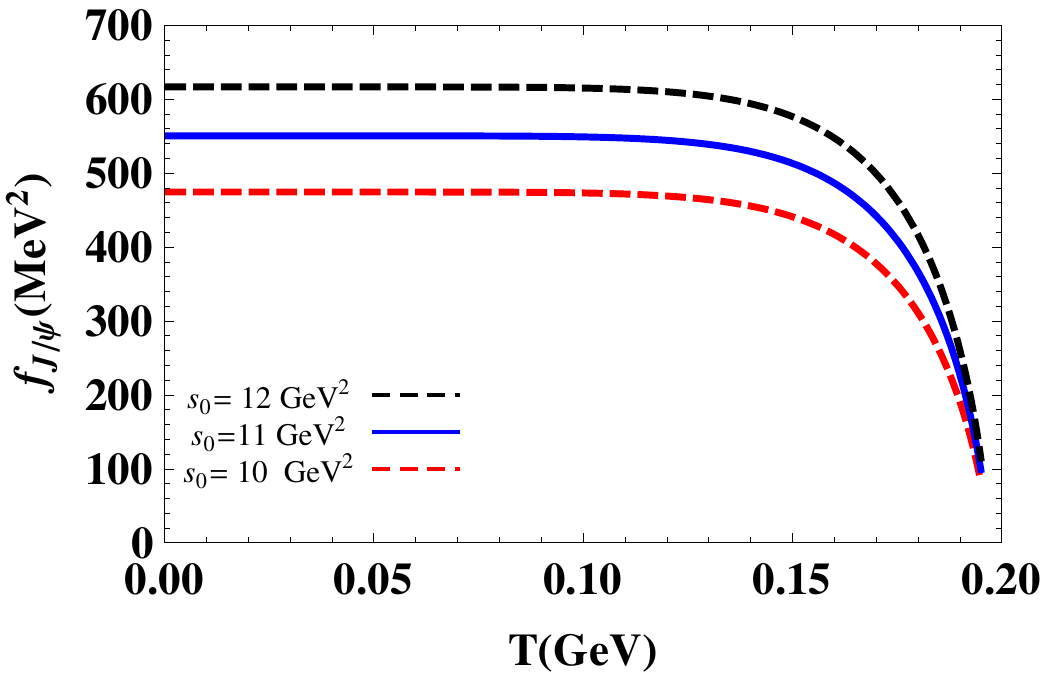}
\includegraphics[width=8cm]{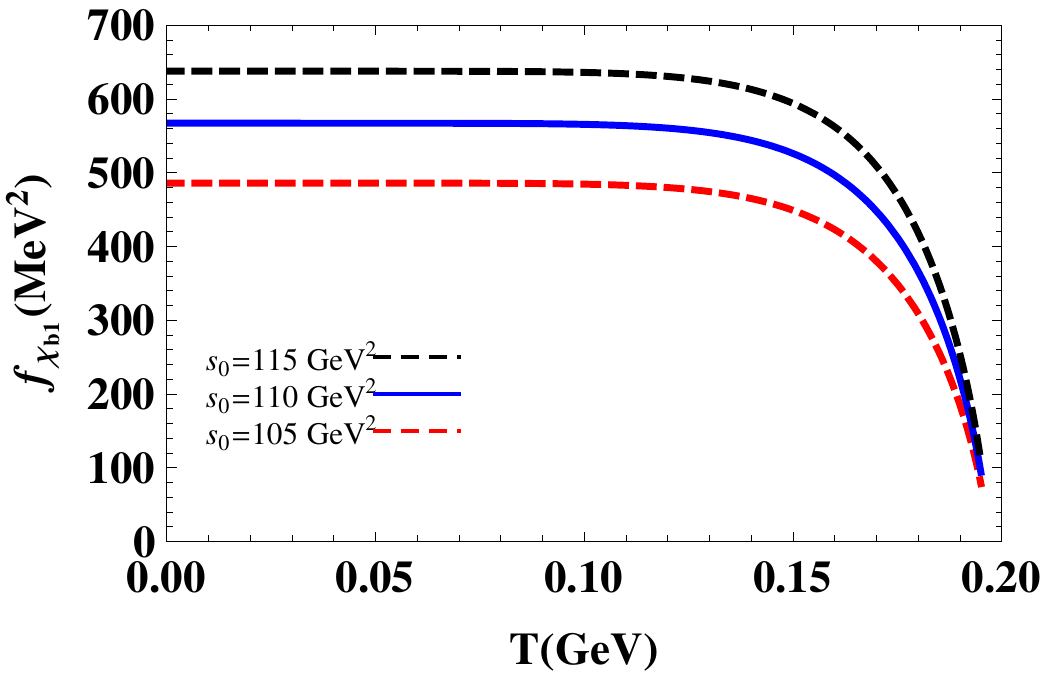}
\includegraphics[width=8cm]{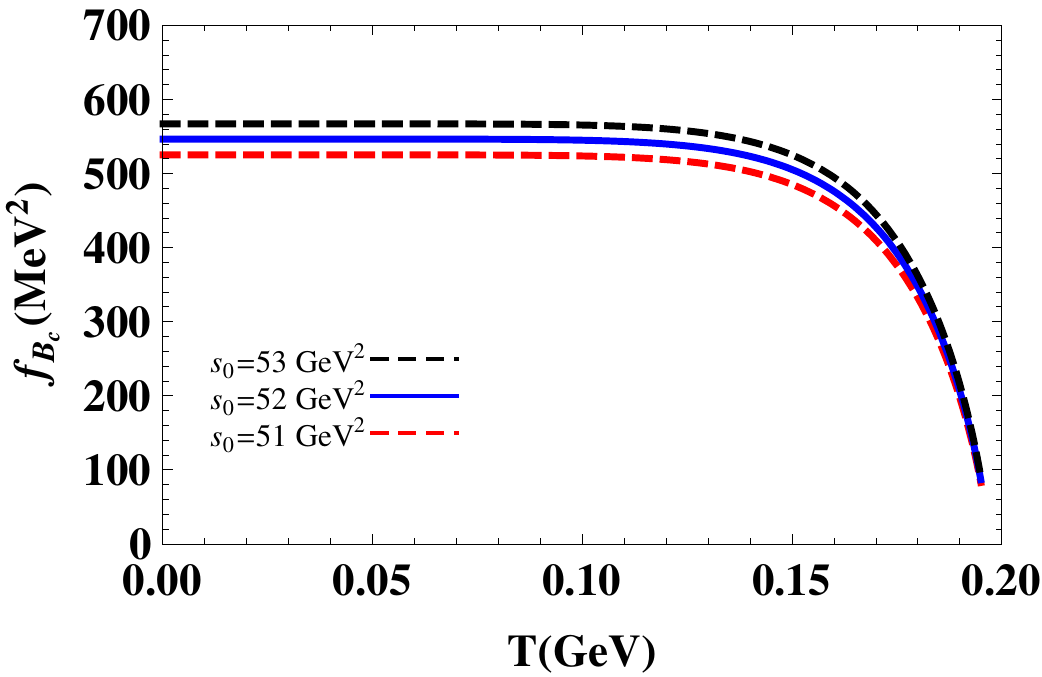}
\includegraphics[width=8cm]{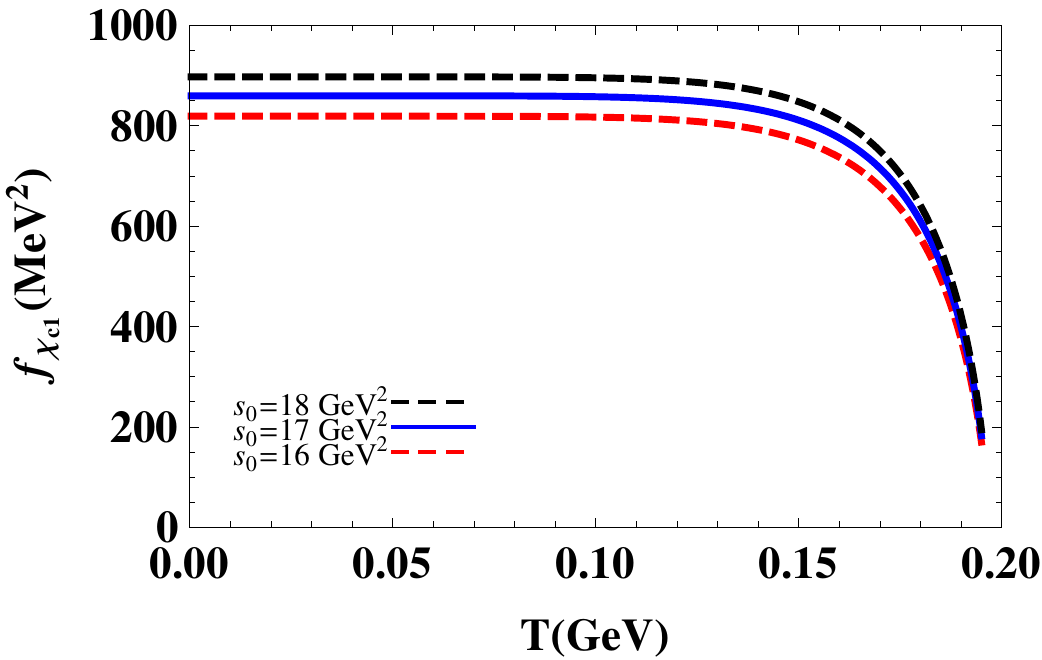}
\end{center}
\caption{Temperature dependencies of vector and axial vector heavy meson decay constants.}
\label{fig6}
\end{figure}

The similar calculations are conducted for the leptonic decay constants of these mesons. The comparison of the results at $T=0$ is given in Table \ref{tab:decays}. In Fig \ref{fig5}, the behaviour of the decay constants in the working region of the Borel mass parameters are presented. Again, we see that the decay constants are stable for the selected $M^2$ intervals. For the decay constant behaviours at finite temperature, Fig \ref{fig6} is presented. In this figure we see that after $T \simeq 150MeV$, a dramatic decrease is observed and around the critical temperature, the decay constants become very close to zero.
 The quoted errors in the results are due to the variations of the Borel parameters as well as of the continuum thresholds and the errors in the input parameters such as quark masses.

\begin{table}[h]
\renewcommand{\arraystretch}{1.5}
\addtolength{\arraycolsep}{3pt}
$$
\begin{array}{|c|c|c|c|c|c|c|}
\hline 
  ~ & f_{\Upsilon}(MeV) &  f_{B_{c}^{1^-}}(MeV) & f_{J/\psi}(MeV) & f_{\chi_{b1}}(MeV) & f_{B_{c}^{1^+}}(MeV)& f_{\chi_{c1}}(MeV) \\
\hline
  Present\ work &  725\pm100   &  375\pm30   &  550\pm65  & 567\pm85 & 546\pm25 & 860\pm40 \\
 \hline
 Experimental \hbox{\cite{Kiselev,Lakhina}} &  708\pm8  &  -  &  409\pm15  &  - & - & - \\
  \hline 
 Thermal\ QCDSR \hbox{\cite{ref12,Axial_quarkonia}} & 746\pm62 & - & 481\pm36 & 240\pm12 & - & 344\pm27 \\
 \hline
 QCDSR\ at\ vacuum \hbox{\cite{WangBc}} & - & 415\pm31 & - & - & 373\pm25 & - \\
 \hline
 Lattice \hbox{\cite{Kiselev,Lakhina}} & - & - & 399\pm4 & - & - & - \\
 \hline
 Non-rel.\ Quark\ Model \hbox{\cite{Lakhina}} & 716 & - & 423 & - & - & - \\
 \hline
 Rel.\ Bethe-Salpeter\ Method \hbox{\cite{Wang_betheMethod}} & 498\pm20 & 418\pm24 & 469\pm28 & - & 160 & - \\
  \hline
\end{array}
$$
\caption{The decay constants of vector and axial vector heavy-heavy mesons at $T=0$ from different theoretical approaches and experimental data.} \label{tab:decays}
\renewcommand{\arraystretch}{1}
\addtolength{\arraycolsep}{-1.0pt}
\end{table}

\section{Conclusion}

Understanding the temperature dependence of hadron properties has a significant role on interpreting experiments focusing on hot and dense QCD matter, such as ALICE at CERN. In order to have a full picture of the hadronic Universe, theoretical studies on thermal behaviours of the observables are crucial. In this article, the mass spectrum of the heavy vector and axial vector mesons as well as the decay constants are studied. Among these mesons, quarkonia spectrum is re-calculated and is compared with previous studies. The results are in good agreement. However, the masses and the decay constants of vector and axial vector $B_c$ mesons at finite temperature were not been calculated up to now. At the critical temperature, the masses are decreased about $3\%$, $5\%$ and $14\%$ for $\Upsilon$, $B_{c}$ and $J/\psi$vector mesons; $6\%$, $7\%$ and $22\%$ for $\chi_{b1}$, $B_{c}$ and $\chi_{c1}$ axial vector mesons, respectively. 

 One of the aims of this study is to obtain the masses and the decay constants of the heavy mesons as the functions of temperature, in order to use them in a future study on the form factors at finite temperature. The results of this study will be fitted into analytical functions and will be used to calculate the strong form factors and also strong coupling constants at $T=0$, as conducted in \cite{Yazici}. Besides, the behaviour of the masses and the decay constants at finite temperature can be checked in future experiments.

\end{document}